\documentclass[%
 reprint,
 amsmath,amssymb,
 aps,
]{revtex4-2}

\pdfoutput=1

\usepackage{graphicx}
\usepackage{dcolumn}
\usepackage{bm}

\usepackage[hidelinks]{hyperref}
\usepackage[hyphenbreaks]{breakurl}
\usepackage{color,array}
\usepackage{graphicx}
\usepackage{CJKutf8}
\usepackage{amsmath}
\usepackage{placeins}
\usepackage{xcolor}
\usepackage{orcidlink}
\usepackage[english]{babel}
\usepackage{subcaption}
\usepackage{color,soul}

\begin{document}


\title{Design of superparamagnetic nanoparticle-materials for high-frequency inductor cores}

\author{Mathias Zambach$^1$\orcidlink{0000-0001-9771-4613}}
\author{Ziwei Ouyang$^2$\orcidlink{0000-0001-7046-9224}}
\author{Matti Knaapila$^{1,3}$\orcidlink{0000-0002-4114-9798}}
\author{Marco Beleggia$^{4,5}$\orcidlink{0000-0002-2888-1888}}
\author{Cathrine Frandsen$^1$\orcidlink{0000-0001-5006-924X}}\email{fraca@fysik.dtu.dk}

\affiliation{$^1$DTU Physics, Technical University of Denmark, 2800 Kgs. Lyngby, Denmark}
\affiliation{$^2$DTU Electro, Technical University of Denmark, 2800 Kgs. Lyngby, Denmark}
\affiliation{$^3$Department of Physics, Norwegian University of Science and Technology, 7491 Trondheim, Norway}
\affiliation{$^{4}$Department of Physics, University of Modena and Reggio Emilia, 41125 Modena, Italy}
\affiliation{$^{5}$DTU Nanolab, Technical University of Denmark, 2800 Kgs. Lyngby, Denmark}

\date{\today}

\begin{abstract}
The progress in the semiconductor industry has resulted in great demand for high-frequency magnetic materials that can be applied to micro-fabricated inductor cores. Nanocomposite materials, containing magnetic nanoparticles in a non-conducting matrix, may provide a solution for materials with high susceptibility or permeability and low power loss in the MHz regime, where traditional ferrites fail in performance. Here, we present a design guide for usage of magnetic nanoparticles in such materials. With statistical mechanics methods, we derive the magnetic susceptibility of nanoparticles in case of uniaxial or cubic anisotropy, as a function of particle size and applied field direction. We also investigate the role of particle shape and interactions. Using the derived susceptibilities, with inductor core applications in mind, we show that near-spherical particles of materials with high saturation magnetization and low magnetic anisotropy, such as FeNi$_3$, are optimal. Additionally, we find that the particle size shall be as large as possible while maintaining superparamagnetic behaviour at the relevant operation frequency. Based on this, we predict that high particle susceptibilities of $>$700 (/$>$1500) are possible for randomly oriented (/uniaxially aligned) 20$\pm$1 nm diameter FeNi$_3$ particles, together with high-frequency stability, shown by low out-of-phase component at 2 MHz. Our findings imply that materials containing nanoparticles have the potential to be tuned to outperform state-of-the-art ferrite inductor core materials at MHz-frequencies.

\end{abstract}


\maketitle

\section{Introduction}\label{introduction}
Magnetic components, such as inductors and transformers, are essential for power electronics in many handheld devices. However, realisation of efficient micro-inductors is challenging, as inductance scales with size \cite{PSMA_yawger2022a,araghchini2013a}. Although a decrease in inductance can be counteracted by higher operating frequencies, magnetic materials available today become inefficient and heat up rapidly at elevated frequencies due to eddy-current losses \cite{TDK,Fairite,petrecca2019a}. Miniaturisation of magnetic components in electronics is therefore limited by the performance of soft magnetic materials \cite{hurley2018a}. Several roadmaps for power electronics identify the lack of suitable magnetic materials as one of the major obstacles on the way to achieving smaller, faster, greener, and more efficient electronic devices \cite{PSMA_yawger2022a,hurley2018a}. 

\begin{figure}[bp!]
    \centering
    \includegraphics[width=20pc]{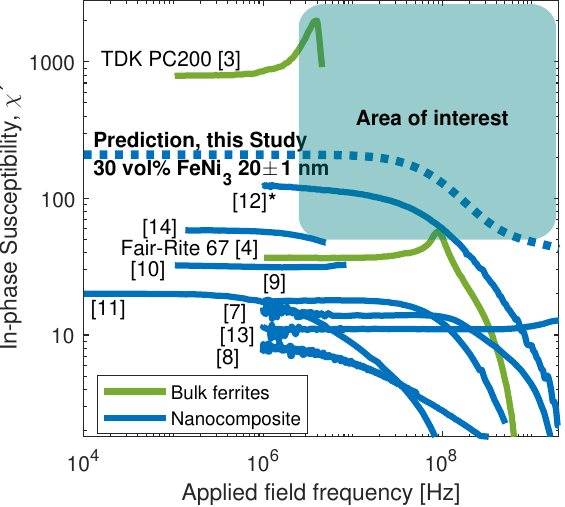}
    \caption{In-phase susceptibility ($\chi'$) for bulk ferrites and nanocomposites {including} the {theoretical} prediction from this study. Reported relative in-phase permeabilities $\mu_{\textrm{r}}'$ have been converted as $\chi'$=$\mu_{\textrm{r}}'-1$. *: Particle (not sample) susceptibility.}
    \label{fig:ACDiscussion}
\end{figure}

The challenge lies in achieving sufficiently high magnetic susceptibility while avoiding significant power losses from eddy currents and magnetic hysteresis at high operation frequencies. To this end, for a magnetic material to be used as a core for micro-inductors, its in-phase susceptibility $\chi'$ needs to be above 50-100 (the shaded region shown in Fig. \ref{fig:ACDiscussion}), while losses should remain below 200 mW/cm$^3$ at an operating frequency of 2 MHz and an induced flux density of 30 mT \cite{TDK}. If losses are reduced below 20 mW/cm$^3$, a $\chi'$-range of 20-50 would be acceptable \cite{PSMA_yawger2022a,hurley2018a}.

For today's power converters operating at up to about 0.5 MHz the typical core materials are sintered, fine-grained ceramics of ferrites (MnZnNiFe-oxides), with poor conductivity that limits eddy current losses. The properties of the ferrites vary with their grain sizes and material composition.
For example, the commonly used ferrite TDK PC200 has a bulk susceptibility of $\sim$800 (shown in Fig. \ref{fig:ACDiscussion}) but only up $\sim$1-2 MHz, where eddy current losses eventually increases drastically and the susceptibility drops \cite{TDK}. New ferrite materials, like Fair-Rite 67 (also shown in Fig. \ref{fig:ACDiscussion}), enable higher operation frequency, but have lower susceptibilities \cite{Fairite}. So far, attempts to make inductor core materials to cover applications in the high-frequency high-susceptibility regime (the shaded area of interest in Fig. \ref{fig:ACDiscussion}) have not been successful \cite{petrecca2019a} and therefore the field of power electronics lacks alternative material solutions.

Magnetic composites using nanoparticles have recently emerged as research materials \cite{FeNi3_Article_Lu,yun2014a,yun2016a,yatsugi2019a,liu2005a,kura2012a,kura2014a,yang2018a,kin2016a,garnero2019a,hasegawa2009a,rowe2015a}, but their potential as inductor materials has not yet been fully explored. The two main attractive reasons for using magnetic nanoparticles to obtain high susceptibility and low power losses are: 1) the nanoparticles, even if metallic, are so small that eddy currents are negligible, and they can be electrically insulated by a non-conductive matrix, 2) the potential high susceptibility and absence of hysteresis within their superparamagnetic regime could be exploited. {Superparamagnetism arises when the magnetic moments of the nanoparticles fluctuate on a faster timescale than the applied field}. This depends on particle size, shape, anisotropy, temperature, applied field frequency, and also {magnetic interparticle} interactions \cite{fock2018a}. 

Several groups have reported susceptibility data for materials containing magnetic nanoparticles ($<100$ nm in diameter), as presented in Fig. \ref{fig:ACDiscussion}. The measured susceptibilities are generally too low to match the area of interest for future power electronics (the shaded area in Fig. \ref{fig:ACDiscussion}). However, it has not been clarified whether composites containing superparamagnetic particles could reach the desired susceptibility in the MHz-range if they were optimized for it. Recently, it has been shown that the susceptibility of single-domain nanoparticles is not limited by their shape (in contrast to soft multi-domain micro-particles) \cite{Zambach2025-DemagPaper}. Hence, there is in principle no upper limit for the effective susceptibility of such nanoparticles, but its actual value will depend on material parameters such as magnetic anisotropy and saturation magnetization as well as particle size and shape.

To evaluate the potential of nanoparticles for use as inductor core materials, we developed a framework for calculating the DC- and AC-susceptibility of single-domain magnetic nanoparticles, both superparamagnetic and blocked. We derive the susceptibility using reported parameter values for specific materials, and considering both uniaxial and cubic anisotropy, combined with the size, shape, and alignment of the nanoparticles. Our theoretical framework reproduces the nanocomposite results shown in Fig. \ref{fig:ACDiscussion}, and based on this predictive power, it provides a way forward to tune and optimize properties and performance to reach the region of interest (Fig. \ref{fig:ACDiscussion}).

\section{Theoretical Framework}\label{article-components}
We consider a stationary single-domain ferromagnetic particle embedded in a solid, non-magnetic matrix. For single-domain particles, atom spins align ferro/ferrimagnetically and rotate coherently such that the magnetic moment of the particle is $m = V M_{\textrm{s}}$, where $V$ is the particle volume and $M_\mathrm{s}$ its saturation magnetization. 

\subsection{Energy considerations}
We define here the various energy terms used later for susceptibility derivations.

\begin{figure}
    \centering
    \begin{subfigure}[t]{0.48\linewidth}
        \centering
        \includegraphics[width=0.7\linewidth]{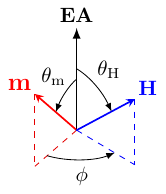}
        \caption{Uniaxial Anisotropy}
        \label{fig:Coor_Uni}
    \end{subfigure}%
    ~ 
    \begin{subfigure}[t]{0.48\linewidth}
        \centering
        \includegraphics[width=\linewidth]{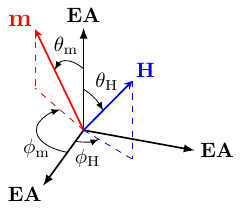}
        \caption{Cubic Anisotropy}
        \label{fig:Coor_Cubic}
    \end{subfigure}
    \caption{Illustration of used coordinate systems and definitions of angles. EA refers to easy axis, $\mathbf{m}$ is the particle moment, and $\mathbf{H}$ is the applied field. (a) Uniaxial anisotropy. (b) Cubic anisotropy with easy axes shown for $K_{\textrm{c}}>0$.}
    \label{fig:Coordinate}
\end{figure}

For uniaxial anisotropy, we choose the polar axis to coincide with the magnetic easy axis, as illustrated in Fig. \ref{fig:Coor_Uni}, the anisotropy energy per particle is then
\begin{equation}
    E_{\textrm{ua}} = K_{\textrm{u}} V \sin^2\theta_{\textrm{m}},
    \label{eq:EAni}
\end{equation}
where $K_{\textrm{u}}$ is the uniaxial anisotropy constant for the material, and $\theta_{\textrm{m}}$ is the angle between the magnetisation and the easy axis of magnetisation, i.e. the polar angle of the magnetisation.
The Zeeman energy is
\begin{equation}
    E_{\textrm{Z}} = -\mu_0 m H \left( \cos\theta_{\textrm{m}}\cos\theta_{\textrm{H}}+\sin\theta_{\textrm{m}}\sin\theta_{\textrm{H}}\cos\phi \right),
    \label{eq:EZee}
\end{equation}
where $\theta_{\textrm{H}}$ is the polar angle of the applied field, $H$ is the applied field amplitude, and $\phi = \phi_{\textrm{H}} - \phi_{\textrm{m}}$ is difference between the azimuth angles of the applied field and the magnetisation.

For cubic anisotropy we orient the coordinate system based on the case where the cubic anisotropy constant is positive, K$_c>$ 0, i.e., 3 mutually orthogonal easy axes, see Fig. \ref{fig:Coor_Cubic}{ but consider both K$_c>$ 0 and K$_c<$ 0 in the following derivations}. One easy axis is set to the polar axis and the other two easy axis are along the direction where the azimuth is 0 and $\pi/2$. The anisotropy energy can then be expressed as
\begin{equation}
    E_{\textrm{c}}
    = K_{\textrm{c}} V \sin^2\theta_{\textrm{m}} \left( \cos^2\theta_{\textrm{m}} + \sin^2\theta_{\textrm{m}}\sin^2\phi_{\textrm{m}}\cos^2\phi_{\textrm{m}} \right),
    \label{eq:EAniCubic}
\end{equation} 
where $K_{\textrm{c}}$ is the cubic anisotropy constant. The Zeeman energy for the cubic case is
\begin{equation}
    E_{\textrm{Z}} = -\mu_0 V M_s H \begin{bmatrix}
    \sin\theta_{\textrm{m}}\cos\phi_{\textrm{m}}\\
    \sin\theta_{\textrm{m}}\sin\phi_{\textrm{m}}\\
    \cos\theta_{\textrm{m}}
    \end{bmatrix}\cdot
    \begin{bmatrix}
    \sin\theta_{\textrm{H}}\cos\phi_{\textrm{H}}\\
    \sin\theta_{\textrm{H}}\sin\phi_{\textrm{H}}\\
    \cos\theta_{\textrm{H}}
    \end{bmatrix}.
\end{equation}
{This takes the same form as} \eqref{eq:EZee} {for the uniaxial case when the vector scalar product is performed. The distinction of the angles are, however, of importance for the angular integrations in} \eqref{eq:partitionfct} {and therefore the vector components are written out here.}

For spheroids, and in general for rotational symmetric, uniformly magnetized bodies, the demagnetising field is $\mathbf{H}_{\textrm{d}} = -\mathbf{N}\mathbf{M}$, where $\mathbf{N}$ is the demagnetisation tensor and $\mathbf{M}$ is the magnetisation. The demagnetisation energy then has the same form as the uniaxial anisotropy energy:
\begin{equation}
    E_{\textrm{H}_\textrm{d}} = -\frac{\mu_0}{2} \int_V \mathbf{M}\cdot\mathbf{H}_{\textrm{d}} \textrm{d}V = K_{\textrm{sh}} V \sin^2\Theta,
    \label{eq:Edemag}
\end{equation}
with an effective shape anisotropy constant, $K_{\textrm{sh}}= \mu_0 M_s^2\left( N_{\textrm{a}} - N_{\textrm{b}} \right)/2$. $N_{\textrm{a}}$ and $N_{\textrm{b}}$ are the demagnetisation factors along the principal spheroid axes. $K_{\textrm{sh}}$ is positive (/negative) for prolate (/oblate) spheroids, respectively, and $\Theta$ is the angle between the magnetic moment and the longer (/shorter) principal spheroid axis. For most soft magnetic materials, shape anisotropy dominates over magneto-crystalline anisotropy if the length difference between the axes is larger than 5-10\%.

Dipolar interaction energy between a pair of particles can be written as
\begin{align}
    E_{\textrm{ia}} &= \frac{\mu_0 V^2 M_s^2}{4\pi r_{\textrm{cc}}^3} \left\{ \begin{bmatrix}
    \sin\theta_{\textrm{m},1}\cos\phi_{\textrm{m},1}\\
    \sin\theta_{\textrm{m},1}\sin\phi_{\textrm{m},1}\\
    \cos\theta_{\textrm{m},1}
    \end{bmatrix}\cdot
    \begin{bmatrix}
    \sin\theta_{\textrm{m},2}\cos\phi_{\textrm{m},2}\\
    \sin\theta_{\textrm{m},2}\sin\phi_{\textrm{m},2}\\
    \cos\theta_{\textrm{m},2}
    \end{bmatrix} \right.\nonumber\\
    & - \frac{3}{r_{\textrm{cc}}^2} \left(\begin{bmatrix}
    \sin\theta_{\textrm{m},1}\cos\phi_{\textrm{m},1}\\
    \sin\theta_{\textrm{m},1}\sin\phi_{\textrm{m},1}\\
    \cos\theta_{\textrm{m},1}
    \end{bmatrix}\cdot
    \begin{bmatrix}
    r_{\textrm{cc},x}\\
    r_{\textrm{cc},y}\\
    r_{\textrm{cc},z}
    \end{bmatrix}\right)\nonumber\\
    & \left.\left(\begin{bmatrix}
    \sin\theta_{\textrm{m},2}\cos\phi_{\textrm{m},2}\\
    \sin\theta_{\textrm{m},2}\sin\phi_{\textrm{m},2}\\
    \cos\theta_{\textrm{m},2}
    \end{bmatrix}\cdot
    \begin{bmatrix}
    r_{\textrm{cc},x}\\
    r_{\textrm{cc},y}\\
    r_{\textrm{cc},z}
    \end{bmatrix}\right)
    \right\}.
\end{align}
Here $r_{\textrm{cc}}$ is the center-center distance of the two spheres, $x,y,z$ denotes its Cartesian components, and the subscript $1,2$ for the angles denote the particle number.

\subsection{Susceptibility calculations}
We derive the susceptibility of single domain particles for the two limiting cases: blocked particles with no thermal agitation, and superparamagnetic particles with their magnetic moments in rapid thermal fluctuation. The characteristic timescale, $\tau$, for relaxation of the moment between easy directions, depends exponentially on the anisotropy barrier over the thermal energy $K_{\textrm{u}}V/(k_B T)$.
Magnetic particles can thus be classified depending on the timescale of superparamagnetic relaxation versus the experimental timescale.  

For blocked particles, thermal fluctuations are slower than the experimental timescale. In case of uniaxial anisotropy, the susceptibility depends on the direction of the applied field with respect to the particle easy axis, as described by the Stoner-Wohlfarth model \cite{stoner1948a}. For randomly oriented particles, the orientation averaged particle susceptibility $\langle\chi_\textrm{B}\rangle$ is 
\begin{equation}
    \langle\chi_\textrm{B}\rangle = \langle \cos^2\theta_{\textrm{H}} \rangle\chi_{\textrm{B}}(0) + \langle \sin^2\theta_{\textrm{H}} \rangle \chi_{\textrm{B}}(\pi/2)
    = \frac{\mu_0 M_{\textrm{s}}^2}{3 K_{\textrm{u}}}
    \label{eq:blockChi}
\end{equation}
since $\langle \cos^2\theta_{\textrm{H}} \rangle = 1/3$ and $\langle \sin^2\theta_{\textrm{H}} \rangle = 2/3$ and the blocked susceptibilities are $\chi_{\textrm{B}}(0)=0$ and $\chi_{\textrm{B}}(\pi/2) = \mu_0 M_{\textrm{s}}^2/\left( 2 K_u \right)$ \cite{svedlindh1997a}. Blocked particles with uniaxial anisotropy axis aligned parallel to the applied field feature a square hysteresis loop with coercive field of H$_\textrm{c} = 2 K_\textrm{u}/(\mu_0 M_\textrm{s})$. If the easy axis is perpendicular to the field, instead, no hysteresis is observed. For blocked particles with cubic anisotropy, a hysteresis loop opening is always observed no matter the direction of the applied field with respect to the 3 or 4 easy axes and thus only a small low field susceptibility is expected \cite{joffe1974a,walker1993a,usov1997a}.

For superparamagnetic particles, thermal fluctuations are faster than the experimental timescale. The magnetisation and susceptibility can be found as for a paramagnetic ion. The partition function for a particle can be written as
\begin{equation}
    Z = \int_\Omega \exp\left[ - \frac{E_{\textrm{i}}(\Omega) }{k_B T} \right]\,\textrm{d}\Omega,
    \label{eq:partitionfct}
\end{equation}
where the integral over $\Omega$ indicates integration over all possible energy states $E_{\textrm{i}}(\Omega)$. The relevant single particle energies presented in Eqs. \eqref{eq:EAni}-\eqref{eq:Edemag} depend on moment direction, thus the integral in Eq. \eqref{eq:partitionfct} should be carried out over all possible moment directions: $\int_\Omega \textrm{d}\Omega = \int_0^{2\pi} \textrm{d}\phi_{\textrm{m}} \int_0^\pi \sin\theta_{\textrm{m}} \textrm{d}\theta_{\textrm{m}} $.
To ease notation we define following energy ratios
\begin{equation}
    \epsilon_{\textrm{k}} = \frac{K_{\textrm{u}}V}{k_B T},
    \,\,\,
    \epsilon_{\textrm{H}} =  \frac{\mu_0 V M_{\textrm{s}} H }{k_B T},
    \,\,\,
    \epsilon_{\textrm{M}} = \frac{\mu_0 V M_{\textrm{s}}^2}{k_B T}.
    \label{eq:Epsilon}
\end{equation}
The unit-less component of the mean magnetic moment of a superparamagnetic particle along the direction of the applied field, $\langle m_{\textrm{SPM}}\rangle$, can now be found as
\begin{align}
    \langle m_{\textrm{SPM}}\rangle &= \frac{1}{Z}\,\, \frac{\partial Z} {\partial \epsilon_{\textrm{H}}}.
    \label{eq:meanm}
\end{align}
The mean magnetisation along the direction of the applied field is then $\langle M_{\textrm{SPM}}\rangle = M_{\textrm{s}}\langle m_{\textrm{SPM}}\rangle$.
The susceptibility is found as the derivative of the mean magnetisation with respect to the applied field amplitude
\begin{align}
    \chi_{\textrm{spm}}=M_{\textrm{s}}\frac{\partial\langle m_{\textrm{SPM}}\rangle}{\partial H}.
\end{align}
For the non-interacting, uniaxial anisotropy particle case, with the applied field at an angle $\theta_{\textrm{H}}$ to the easy axis, one finds the superparamagnetic particle susceptibility $\chi_{\textrm{spm}}$ to be
\begin{align}
    \chi_{\textrm{spm}}(\theta_{\textrm{H}}) = \frac{\epsilon_{\textrm{M}}}{2}\left[\sin^2\theta_{\textrm{H}} + {R'}/{R}\left(3\cos^2\theta_{\textrm{H}}-1\right) \right].
    \label{eq:ChiAngle}
\end{align}
with 
\begin{equation}
    R' = \int_0^1 x^2 \exp\left( \epsilon_{\textrm{k}} x^2 \right) \textrm{d}x \quad\textrm{and}\quad R = \int_0^1 \exp\left( \epsilon_{\textrm{k}} x^2 \right) \textrm{d}x.
    \label{eq:RRdiff}
\end{equation}
 
\begin{figure}[t]
\centerline{\includegraphics[width=20pc]{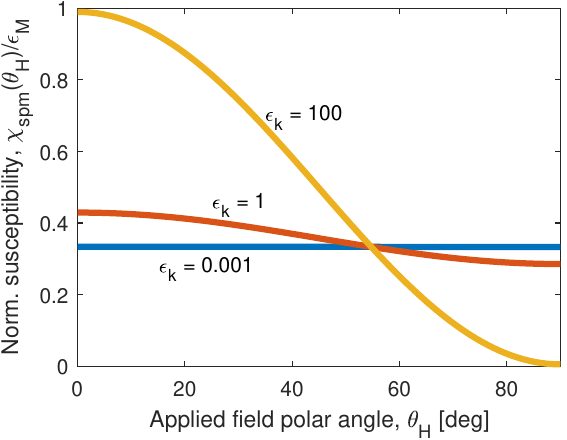}}
\caption{Superparamagnetic particle susceptibility normalised by $\epsilon_{\textrm{M}}$ as function of the angle between applied field and uniaxial anisotropy easy axis, $\theta_{\textrm{M}}$, for {non-interacting} particles with varying anisotropy barrier $\epsilon_{\textrm{k}}$. Based on Eqs. \eqref{eq:ChiAngle}-\eqref{eq:RRdiff}.
}
\label{fig:ChiAngle}
\end{figure}

Figure \ref{fig:ChiAngle} shows a plot of Eq. \eqref{eq:ChiAngle} for different values of $\epsilon_{\textrm{k}}$. The susceptibility ranges from $\epsilon_{\textrm{M}}$ to 0 in case of large anisotropy ($\epsilon_{\textrm{k}}\gg1$). For low anisotropy ($\epsilon_{\textrm{k}}\ll1$) $\chi_\mathrm{spm}$ goes towards the random case value $\epsilon_{\textrm{M}}/3$ for all $\theta_{\textrm{H}}$. {Special cases of} \eqref{eq:ChiAngle} {are found in}  \cite{ludwig2017a,raikher1974a,shliomis1993a,eisenstein1977a,kalmykov2000a,elfimova2019a}.

The orientation average of eq. \eqref{eq:ChiAngle} is the well known superparamagnetic susceptibility
\begin{align}
    \langle \chi_{\textrm{spm}} \rangle = \frac{\epsilon_{\textrm{M}}}{3} = \frac{\mu_0 V M_{\textrm{s}}^2}{3 k_B T}.
    \label{eq:ChiRandom}
\end{align}

For a non-interacting particle with cubic anisotropy we find no dependence of the susceptibility on easy axis direction with respect to the applied field (i.e. on the angles $\theta_{\textrm{H}}$ and $\phi_{\textrm{H}}$). The initial susceptibility for the cubic anisotropy case is always the same as for the randomly oriented uniaxial anisotropy case in the superparamagnetic regime (Eq. \eqref{eq:ChiRandom}), no matter the strength of the anisotropy {constant}. This can be explained by the symmetric distribution of energy minima along the 3 or 4 easy axes, which does not change the probability of the moment to point along any specific direction, contrary to the uniaxial case where there is 1 direction (or plane) that is statistically more likely.

Dipolar interactions can be included in the above derivations. For the two particle case, assuming low anisotropy particles ($\epsilon_{\textrm{k}}\ll1$), we find that the susceptibility per particle in case of interactions, $\chi_{\textrm{ia}}$, when the two particles form a dimer oriented parallel or perpendicular to the field direction, is, respectively 
\begin{equation}
    \chi_{\textrm{ia,}||} = \frac{\epsilon_{\textrm{M}}}{3} \left( 1 + \frac{2}{96}\frac{\epsilon_{\textrm{M}} V}{\pi r_{\textrm{cc}}^3} \right),
    \,\,
    \chi_{\textrm{ia,}\perp} = \frac{\epsilon_{\textrm{M}}}{3} \left( 1 - \frac{1}{96}\frac{\epsilon_{\textrm{M}} V}{\pi r_{\textrm{cc}}^3} \right).
    \label{eq:Interaction1}
\end{equation}
Here $r_{\textrm{cc}}$ is the center-center distance of the two particles and $\epsilon_{\textrm{M}}/3$ would be the susceptibility without interaction.

The dependence of the susceptibility on applied field frequency and the crossover from superparamagnetic to blocked regime can be described and analysed via the concept of AC-susceptibility \cite{svedlindh1997a,ludwig2017a}. At small $\epsilon_\textrm{H}$, where the anisotropy barrier is not affected significantly by the applied field, the AC-susceptibility for a sinusoidal field with angular frequency $\omega$, $\tilde{\chi}(\omega)$, is connected to the superparamagnetic particle susceptibility $\chi_{\textrm{spm}}$ and the superparamagnetic relaxation time $\tau$ as in the Debye model 
\begin{equation}
    \tilde{\chi}(\omega) = \frac{\chi_{\textrm{spm}}}{1 + i \omega \tau} = \chi'(\omega) + i \chi''(\omega),
    \label{eq:Debye}
\end{equation}
with in- and out-of-phase components $\chi'(\omega)$ and $\chi''(\omega)$ \cite{carrey2011a}.

For uniaxial anisotropy particles $\tau$ depends on the direction of the applied field with respect to the easy axis \cite{raikher1974a,shliomis1993a,svedlindh1997a,ludwig2017a}. The magnetic moment relaxation time perpendicular to the anisotropy axis $\tau_\perp$ is assumed to be short, on the timescale of the attempt time $\tau_0\approx 10^{-11}-10^{-9}\,$s \cite{fock2018a}. The relaxation time over the anisotropy barrier $\tau_\parallel$, (parallel to the anisotropy axis), is slower and often described by Arrhenius-type expressions. The AC-susceptibility for randomly oriented uniaxial anisotropy particles, $\langle\tilde{\chi}_{\textrm{spm}}(\omega)\rangle$, can thus be expressed by the weighted sum of the AC-susceptibility parallel and perpendicular with the easy axis:
\begin{equation}
    \langle\tilde{\chi}_{\textrm{spm}}(\omega)\rangle = \frac{1}{3}\left[ \frac{\chi_{\textrm{spm}}(0)}{1 + i\omega\tau_{\parallel} }  + 2\frac{\chi_{\textrm{spm}}(\pi/2)}{1 + i\omega\tau_{\perp} } \right],
    \label{eq:ACSusc}
\end{equation}
with $\chi_{\textrm{spm}}(0)$ and $\chi_{\textrm{spm}}(\pi/2)$ from Eq. \eqref{eq:ChiAngle}. The relaxation times $\tau_\parallel$ and $\tau_\perp$ are found as \cite{raikher1974a,shliomis1993a}
\begin{subequations}
\begin{align}
    \tau_\parallel &= \begin{cases}\tau_0\,2 R'/(R-R')
    \\
    \tau_0 \sqrt{\pi}\exp(\epsilon_{\textrm{k}})/(2 \epsilon_{\textrm{k}}^{3/2})
    \end{cases}
    \quad &\begin{array}{l}
        \textrm{for } \epsilon_{\textrm{k}}\leq2,\\[3pt]
        \textrm{for } \epsilon_{\textrm{k}} > 2,
    \end{array}\\
    \tau_\perp &= \tau_0\, 2 \left( R - R' \right)/\left( R + R' \right) 
    \quad &\textrm{for all } \epsilon_{\textrm{k}}.
\end{align}
\label{eq:relaxDir}
\end{subequations}
The out-of-phase component gives us an estimate of the hysteresis loss per particle volume by $P_{\textrm{H}}=\omega \mu_0 H^2 \chi''(\omega) /2 $ for small applied field amplitudes $H$. For low anisotropy particles the out-of-phase component of the AC-susceptibility is overestimated by the treatment presented above, and should be regarded as an upper bound \cite{carrey2011a}.

Expressions similar to the parallel case of \eqref{eq:relaxDir} exist for cubic anisotropy, but with a lower effective anisotropy constant $K_{\textrm{eff}}$ of $K_{\textrm{eff}} = K_{\textrm{c}}/4$ for $K_{\textrm{c}}>0$ and $K_{\textrm{eff}} = K_{\textrm{c}}/12$ for $K_{\textrm{c}}<0$ \cite{eisenstein1977a,smith1976a,kalmykov2000a,kalmykov2002a}.

The formalism introduced above enables the estimate of the maximum susceptibilities of nanoparticles, which is the topic of the next section. We consider candidate materials to be used with applied field at frequencies of up to 10 MHz.

\section{Particle optimization for nanocomposites}
\begin{figure}[t]
\centerline{\includegraphics[width=20pc]{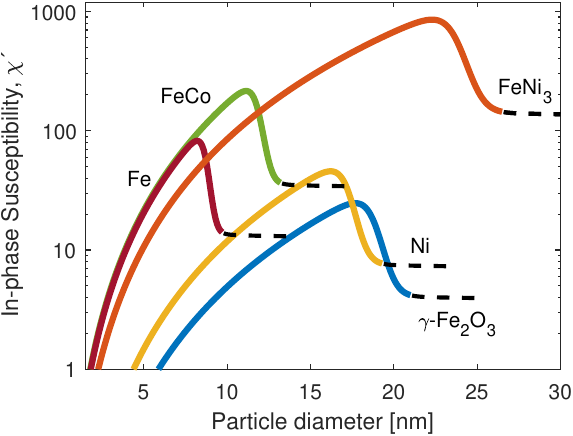}}
\caption{Average per particle in-phase susceptibility for randomly oriented, {non-interacting}, spherical particles in 10 MHz sinusoidal fields as a function of particle size. Calculations are based on Eqs. \eqref{eq:ChiAngle} and \eqref{eq:ACSusc}-\eqref{eq:relaxDir}, $\tau_0=10$ ns, $T=298$ K, and material parameters; Maghemite ($\gamma$-Fe$_2$O$_3$): $K_\textrm{u}=10$ kJ/m$^3$, $M_\textrm{s}=303$ kA/m, 
Ni: $K_\textrm{u}=13$ kJ/m$^3$, $M_\textrm{s}=470$ kA/m,
Fe: $K_\textrm{u}=100$ kJ/m$^3$, $M_\textrm{s}=1750$ kA/m,
FeCo: $K_\textrm{u}=40$ kJ/m$^3$, $M_\textrm{s}=1790$ kA/m,
FeNi$_3$: $K_\textrm{u}=5$ kJ/m$^3$, $M_\textrm{s}=1260$ kA/m. Blocked regime indicated by black, dashed lines.
}
\label{fig:ChiRandom}
\end{figure}

Figure \ref{fig:ChiRandom} shows the effective in-phase susceptibility of magnetic nanoparticles as function of size, from the superparamagnetic regime to the blocked regime, for randomly oriented, {non-interacting}, spherical nanoparticles. We have considered a field frequency of 10 MHz and the particle susceptibilities are calculated from Eqs. \eqref{eq:ChiAngle} and \eqref{eq:ACSusc}-\eqref{eq:relaxDir}. Particle materials used are maghemite ($\gamma$-Fe$_2$O$_3$), Fe, Ni, FeNi$_3$, and FeCo. {Despite several of these materials posses cubic crystallinity,} effective uniaxial anisotropy and saturation magnetization values from the literature for nanoparticle materials have been adopted \cite{rivas2022a,bohra2021a,boedker1994a,lacroix2009a,kumari2021a}. The assumed values for ${M}_{\textrm{s}}$ and ${K}_{\textrm{u}}$ are given in the figure caption. 
We have used values reported for 3-10 nm sized particles (21-50 nm for FeNi$_3$) and not considered the possible dependence of saturation magnetisation and effective anisotropy constant on synthesis routine and particle size. 

Figure \ref{fig:ChiRandom} shows that the particle susceptibility increases with particle size for small particles in the superparamagnetic regime. When the particles reach sizes where their superparamagnetic relaxation time is comparable to the timescale of the applied field, the susceptibility peaks. For larger particles, the susceptibility declines towards the blocked value, which depends on particle saturation magnetisation and anisotropy, but not on the particle size, (see Eq. \eqref{eq:blockChi}). Fig. \ref{fig:ChiRandom} illustrates how both FeCo and FeNi$_3$ particles feature  effective particle susceptibilities above 100 in the superparamagnetic region. 

Our results indicate that nanoparticles with high saturation magnetization and low anisotropy have the largest susceptibility, as their magnetic moments are larger and their transition to the blocked regime occurs at larger diameters. The superparamagnetic region towards the susceptibility peak is of particular interest for applications due to the potential combination of high-susceptibility and limited hysteresis loss. Hence, the optimal size for a given material is on the ascending part of the susceptibility curve, before reaching the maximum of the susceptibility.
Based on Fig. \ref{fig:ChiRandom}, we suggest the use of $\sim$9 nm FeCo particles or $\sim$20 nm FeNi$_3$ particles for high-susceptibility applications such as micro-inductor core materials. 

The susceptibility of around 130 for the blocked FeNi$_3$ particles, as seen in Fig. \ref{fig:ChiRandom}, could in principle be of interest, as it may be sufficient for use in micro-inductors. However, blocked particles require precise alignment as only the $\theta_{\textrm{H}}=\pi/2$ case shows no hysteresis losses.
\begin{figure}[t]
\centerline{\includegraphics[width=20.5pc]{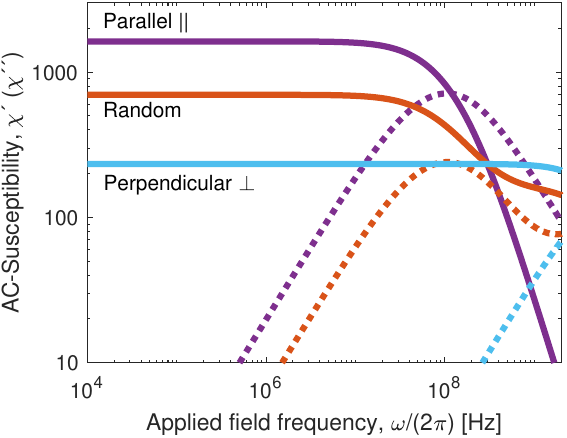}}
\caption{In-phase (solid) and out-of-phase (dashed) susceptibility of {non-interacting} 20$\pm$1 nm diameter (log-normal distributed) FeNi particles with easy axis oriented randomly, parallel or perpendicular to applied field axis as function of applied field frequency. Calculations are based on \eqref{eq:ACSusc}-\eqref{eq:relaxDir}, $\tau_0=10$ ns, $T=298$ K, effective uniaxial anisotropy of $K_\textrm{u}=5$ kJ/m$^3$ and $M_\textrm{s} = 1260$ kA/m. Parallel and perpendicular alignment refer to $\theta_{\textrm{H}} = 0$ and $\pi/2$ respectively.}
\label{fig:ACChiAligned}
\vspace*{-6pt}
\end{figure}

Eqs. \eqref{eq:ACSusc}-\eqref{eq:relaxDir} reveal that also the dynamic susceptibility depends on the direction of the particle with respect to the applied field. Fig. \ref{fig:ACChiAligned} shows the frequency dependence of the in- and out-of-phase susceptibility for uniaxial anisotropy FeNi$_3$ particles with log-normal distributed diameters of 20$\pm$1 nm for cases with anisotropy axes parallel, perpendicularly and randomly oriented with respect to the applied field. In the frequency range of 10 kHz to 10 MHz the 20 nm FeNi particles aligned with their easy axis parallel to the applied field have an in-phase susceptibility of $\sim$1600, i.e., more than twice the value for the randomly oriented case (700) and more than six times the perpendicular "hard-axis" case (230). At around 100 MHz the in-phase part drops and the out-of-phase component peaks for the parallel aligned and random cases. This is due to $\omega \approx 1/\tau_\parallel$ in this frequency range. For the random case the susceptibility drops towards the random orientation blocked susceptibility of Eq. \eqref{eq:blockChi}. For the perpendicular case we observe a completely flat susceptibility up to the GHz regime where $\omega \approx 1/\tau_\perp$, and we find that the susceptibility does not depend on particle size, with values close to the blocked, aligned case of Eq. \eqref{eq:blockChi}. Alignment of particles by easy/hard axis thus allows for tuning of the magnetic properties. 
\begin{figure}[t]
\centerline{\includegraphics[width=20pc]{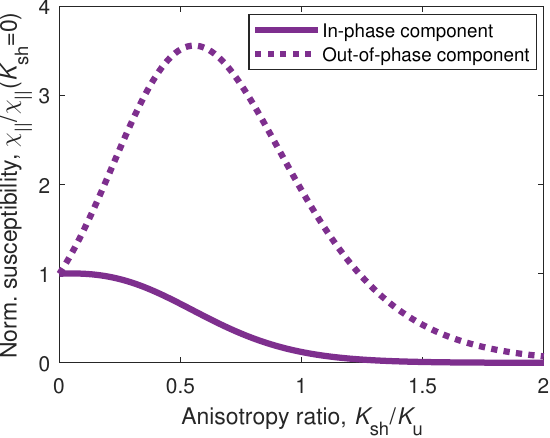}}
\caption{Normalised in-phase (solid) and out-of-phase (dashed) susceptibility as function of shape anisotropy for {non-interacting} 20$\pm$1 nm FeNi$_3$ particles aligned parallel to the applied field. Material values and applied field frequency are as in figure \ref{fig:ChiRandom}, and the effective anisotropy is taken as the sum of a uniaxial anisotropy and the shape anisotropy $K_\textrm{eff} = K_\textrm{u} + K_\textrm{sh}$.}
\label{fig:Shape}
\end{figure}

Even though shape anisotropy can increase susceptibility for particles with their easy axis aligned to the applied field, (see Eq. \eqref{eq:ChiAngle} and Fig. \ref{fig:ChiAngle}), the anisotropy enhancement will increase relaxation time for such particles and thus the out-of-phase component, corresponding to power losses due to magnetic hysteresis. This is illustrated in Fig. \ref{fig:Shape}, where the normalised in- and out-of-phase components of the susceptibility are shown for 20$\pm$1 nm FeNi$_3$ particles aligned with easy-axes along the applied field. We have assumed that shape and magneto-crystalline anisotropy are in the same direction, i.e. $K_{\textrm{eff}} = K_{\textrm{u}} + K_{\textrm{sh}}$. Fig. \ref{fig:Shape} shows that the in-phase component of the susceptibility is relatively insensitive to a shape anisotropy increase ($<$0.3 \% at $K_{\textrm{sh}} = 0.05 K_{\textrm{u}}$), while the increase in out-of-phase component is substantial ($>$22 \% at $K_{\textrm{sh}} = 0.05 K_{\textrm{u}}$). We also observe that for larger shape anisotropy ($K_{\textrm{sh}}>0.6 K_{\textrm{u}}$) the particle becomes blocked.
Particles smaller than the blocking diameter at a given frequency will show a more significant increase in in-phase susceptibility. However, the connected increase of the out-of-phase susceptibility will be similar to the case in Fig. \ref{fig:Shape}. These results on FeNi$_3$ particles lead us to conclude that spherical particles are the most optimal, as any shape anisotropy present increases the out-of-phase component of the susceptibility due to increased superparamagnetic relaxation time, resulting in larger losses per induced magnetisation as discussed in the next section.

Hysteresis losses per particle can be calculated from the out-of-phase component of the susceptibility [Eqs. \eqref{eq:ChiAngle},\eqref{eq:ACSusc} and \eqref{eq:relaxDir}]. For nanocomposites containing superparamagnetic particles we assume a linear dependence of nanocomposite susceptibility on the volume fraction and particle susceptibility. Therefore the applied field strength $H$ needed to obtain a certain magnetic flux in the material varies between materials. Hence, for a given application with a required flux density, the loss might be larger for a low susceptibility material than for high susceptibility materials due to the need for larger applied field. Therefore, assuming no interaction and same particle characteristics, the losses will increase with decreasing volume fraction of particles when the nanocomposites generate equal flux density.
\begin{table}[b]
\caption{Calculated susceptibility and losses of nanocomposites containing 30 vol\% spherical FeNi$_3$ particles with 3 different easy axis arrangements. Calculations are based on $\omega/(2\pi) = 2$ MHz, using Eqs. Eqs. \eqref{eq:ChiAngle},\eqref{eq:ACSusc} and \eqref{eq:relaxDir} with an applied field amplitude $H$ such that 30 mT flux density is achieved in the composite, $\tau_0=10$ ns, $T=298$ K, material parameters as in Fig. \ref{fig:ChiRandom}, and log-normal distributed particle diameters of 20$\pm$1 nm.}
\label{table}
\centering
\begin{tabular}{c|c|c|c|c} 
\parbox{5pc}{\centering Material and \newline orientation} & $\chi'_{\textrm{spm}}$ & $\chi''_{\textrm{spm}}$ & \parbox{2.5pc}{\centering $H$ \newline [A/m]} & \parbox{4.5pc}{\centering Power loss \newline [mW/cm$^3$]} \\
\hline
FeNi$_3$ $||$  & 487 & 11 & 49 & 180 \\
FeNi$_3$ Random & 209 & 3.6 & 114 & 330 \\
FeNi$_3$ $\perp$ & 70 & 0.02 & 341 & 20
\end{tabular}
\vspace*{-6pt}
\label{tab:Loss}
\end{table}

Table \ref{tab:Loss} lists calculated susceptibilities and losses at 2 MHz for nanocomposites of 30 volume \% 20$\pm$1 nm FeNi$_3$ particles with anisotropy axes oriented parallel, perpendicular and randomly with respect to the applied field. The losses are calculated with an applied field amplitude $H$ such that a flux density amplitude of 30 mT is achieved in the nanocomposites. The calculated losses in Table \ref{tab:Loss} indicate that for both the parallel and perpendicularly aligned cases, losses lower than 200 mW/cm$^3$ are possible. The aligned particles have lower losses at a given flux density than the random case due to lower applied field amplitude used to reach the desired flux density. 
The effect of temperature on the power loss at a given flux density can be calculated from the theory derived above. Susceptibility will in general decrease as $1/T$, while the out-of-phase component will decrease further due to lower $\tau_{||}$ and $\tau_{\perp}$. Thus higher field will be needed for higher temperature to reach desired flux density, but the power loss will decrease.
Based on these calculations, the susceptibility and losses of perpendicular and parallel aligned spherical single-domain FeNi$_3$ nanoparticles seem very promising for micro-inductor applications. 

The direct effect of dipolar interaction on the susceptibility can be illustrated by use of equation \eqref{eq:Interaction1}. We find that for a hypothetical FeNi$_3$ 20$\pm$1 nm diameter two-particle system, an increase (/decrease) of susceptibility due to the two particles being oriented along a chain parallel (/perpendicular) to the applied field will be of the order of 0.2 (0.1) \% of the total susceptibility at 30 vol\% (assuming that the volume fraction is $V/r_{\textrm{cc}}^3$). The effect due to the effective field therefore appears rather small, and the change holds true for particles with similar high moment. However, dipole fields of neighbouring particles might affect superparamagnetism \cite{durhuus2025a} and thereby have a larger effect on susceptibility. Hence, we deem direct interaction effects on susceptibility to be of relatively low importance for materials with $<$30 vol\% of particles, as long as the particles are still superparamagnetic. 

The effect of dipolar interactions on the superparamagnetic relaxation is not yet fully described in the literature \cite{durhuus2025a}. It has been suggested that dipolar interactions can slightly lower the effective anisotropy barrier in low volume fraction scenarios\cite{joensson2001a}, while for dense aggregates it is often assumed to drastically increase effective anisotropy, consistently with the observed suppression of superparamagnetism \cite{fock2018a,morup1994a,m2010a,durhuus2025a}.  
From energy considerations we argue that larger particle size with lower volume fraction are preferred, as the increase in uniaxial anisotropy energy is lower than the decrease in dipolar interaction energy, at a fixed nanocomposite susceptibility.

\section{Discussion}\label{discussion}
We have shown how to calculate the susceptibility for magnetic nanoparticle materials. Our theoretical framework provides novel results for particle alignment, cubic anisotropy, and effects from dipolar interactions that are not yet fully accounted for in the literature \cite{elrefai2018a,elfimova2019a}. Moreover, the dependency of particle susceptibility on the applied field direction with respect to the easy axis for uniaxial anisotropy case is interesting for application in power electronics magnetics, as it can be used to tune susceptibility and losses for different materials. We note that the susceptibility of cubic anisotropy particles cannot be tuned by alignment. However, as reported above (Fig. \ref{fig:ChiRandom}) most studies find a mild uniaxial anisotropy in spherical nanoparticles of cubic materials. Since magnetic properties, such as coercive field, are similar for the cubic and random uniaxial cases {when not in the fully blocked state}, it will challenging to distinguish cubic from uniaxial anisotropy when particles are not aligned \cite{usov2012a,ludwig2017a,kura2014a}.

Magnetic nanoparticles have previously been investigated experimentally for use as inductor core materials in the MHz range \cite{FeNi3_Article_Lu,yun2014a,yun2016a,yatsugi2019a,liu2005a,kura2012a,kura2014a,yang2018a,rowe2015a,garnero2019a}, (see Fig.  \ref{fig:ACDiscussion}). Our theoretical framework predicts susceptibility values in quantitative agreement with the general trends expounded by the many experimental reports available in the literature \cite{kura2012a,kura2014a,yun2014a,yun2016a}. This supports the predictive power of our model. The sparse mismatch observed between the model predictions and the data in some situations is ascribed to uncertainties in particle size distributions and material parameters (especially anisotropy and saturation magnetization, which may vary with size). 

The comparison between the experimental in-phase susceptibilities from \cite{FeNi3_Article_Lu,yun2014a,yun2016a,yatsugi2019a,liu2005a,kura2012a,kura2014a,yang2018a} (shown in Fig. \ref{fig:ACDiscussion}) and our model highlights that most studies so far may have not optimized particle materials for the specific applications targeted. Hence, there is room for improvement, and the opportunity to break through the region of interest shown by the shaded area in Fig. \ref{fig:ACDiscussion}. For example, some materials show little or no hysteresis \cite{yun2014a,yun2016a}, which should be a sign of superparamagnetism, but these materials have only relatively low susceptibilities. This is connected to an insufficient saturation magnetisation of the particles (e.g. from oxides) and/or too dilute particle concentrations. For several other materials the nanoparticles were packed too densely or were aggregated \cite{FeNi3_Article_Lu,yatsugi2019a,kura2014a,yang2018a}, which effectively renders the particles blocked and results in large coercive field of several thousand kA/m. Consequently, the resulting materials behave much like those comprising larger particles, i.e. with lower susceptibility and larger coercivity (see for example \cite{liu2005a}). 
One reference reports a high \textit{particle} susceptibility of 122, using relatively densely packed 8 nm Fe particles \cite{kura2012a}, which fits well with the results from Fig. \ref{fig:ChiRandom} given that the resulting particle size is effectively larger than 8 nm. This is promising for micro-inductor applications.

Nanocomposite materials thus show the potential to reach the area of interest of susceptibility above 50 in the frequency range above 2 MHz as indicated in Fig. \ref{fig:ACDiscussion}. Based on our model, we propose that optimised particles in the form of spherical $\sim$9 nm Fe (or FeCo) or $\sim$20 nm FeNi$_3$ particles, have the properties needed to develop inductor cores of the next generation. A difficulty in the applicability of our theoretical framework is that effective particle property values are somewhat sparse. Hence, the reported effective particle values in our evaluation of superparamagnetic particles may not be exact for all given sizes. Moreover, it is not always possible to synthesize particles in the desired size range, optimally monodisperse, and with the specific properties, concentration, or easy axis orientation. A relevant case study would be an uniaxial anisotropy particle accompanied by a small degree of shape anisotropy perpendicular to the uniaxial anisotropy, if they can be synthesized. This combination should produce a lower effective anisotropy barrier \cite{durhuus2024a}.

The alignment of particles with the effective easy axis parallel to the applied field has been investigated experimentally \cite{ludwig2017a,kura2014a}. However, it is not clear whether the observed increase in susceptibility is primarily a result of alignment or of aggregation/chain formation. In fact, any form of clustering might block the particles or, at least, change significantly their dynamic behaviour \cite{durhuus2025a}. To our knowledge, magnetic interactions in the derivation of susceptibilities typically are introduced as an effective field \cite{elfimova2019a} resulting in susceptibility enhancement. On the other hand, Monte-Carlo simulations including the anisotropic dipolar interactions between nanoparticles indicated a susceptibility suppression for increasing volume fraction of particles \cite{chantrell2001a}. Our theoretical framework sheds some light on the matter: it clarifies that for superparamagnetic particles the interactions and their effects depend on particle-assembly orientation with respect to the applied field, as shown in Eq. \eqref{eq:Interaction1} and therefore depends on particle packing. {However, the effect of interparticle interactions is only included in Eq.} \eqref{eq:Interaction1} {and not included in the results presented in Figs. }\ref{fig:ChiAngle}-\ref{fig:Shape}{. The complex effects of dipolar interparticle interactions on susceptibility would be relevant to explore further in future studies.}

Dipolar interactions could potentially have also an effect in the frequency dependence because of the increased effective anisotropy. We note that many materials presented in Fig. \ref{fig:ACDiscussion} are relatively dense materials, and reported susceptibilities of low coercivity materials still seem to have stable susceptibility, although low, even at high frequencies \cite{yun2016a,kura2012a}. The applied field can change the superparamagnetic relaxation, allowing for faster response in applied field than without \cite{Coffey1995a,kalmykov2000a}. This could explain the seemingly high blocking frequency seen for some densely packed materials \cite{kura2012a,yun2016a}.

\section{CONCLUSIONS}\label{conclusion}
We have developed a comprehensive theoretical framework based on statistical mechanics to calculate nanoparticle susceptibility. The framework accounts for the effects of size, shape, anisotropy, saturation magnetization, and inter-particle interactions.

Our exploration of parameter space evidenced how nanoparticles with large saturation magnetisation and low anisotropy are the most suited for inductor cores in power electronics applications. Moreover, we determined that optimal particles should be as large as possible but remain superparamagnetic at the target operation frequency. For this, a narrow particle size distribution is preferable. In relation to particle shape, we found that elongation can increase susceptibility to some degree, but due to the increase in effective anisotropy, it is accompanied by an increase in out-of-phase component, which is associated to magnetic hysteresis losses. Therefore, spherical or near-spherical particle shapes would be most optimal for inductor cores.
Losses per induced magnetisation are minimized for materials where the uniaxial anisotropy axes of the particles are aligned and not randomly distributed. 

As an example of optimised particles, we show that a material with 30 vol\% of spherical 20$\pm$1 nm FeNi$_3$ particles could potentially have a susceptibility above 209 (487 for aligned particles) with lower losses than state-of-the-art ferrite magnetic core materials at $>$2 MHz operation.

The comparison between our model predictions and a broad experimental data set compiled from the literature, shows an excellent agreement, firmly supporting the quantitative predictive power of the model. Sub-optimal susceptibilities for high-frequency inductor core materials ($\chi$´$<$20-50) have been reported for nanocomposites in most experimental studies. This appears as a consequence of a poor choice of materials with low saturation magnetization, imperfect sizing of the particles, and either too dense or too dilute nanocomposites, or too dense aggregates with dipolar interactions suppressing the susceptibility. Our paper demonstrates that better nanocomposite materials are achievable to improve the performance of micro-fabricated inductor cores at high frequencies where bulk ferrite materials such as TDK's PC200 cannot operate.

\begin{acknowledgments}
The authors thank Ron B. Goldfarb for stimulating discussions, and the Independent Research Fund Denmark for financial support (project HiFMag, grant number 9041-00231B).
\end{acknowledgments}

\newpage
\bibliography{apssamp.bib} 

\end{document}